\documentclass{jpsj-suppl}
\usepackage{txfonts} 
\usepackage{graphicx,epsfig}
\usepackage{longtable}
\usepackage{color}
\usepackage[sort&compress,numbers]{natbib}

\usepackage{bm}
\usepackage{capt-of}
\usepackage{relsize}
\usepackage{xcolor}
\usepackage{amssymb}
\usepackage[normalem]{ulem}
\usepackage{mathptmx}
\usepackage[percent]{overpic}

\def\Ga{\Gamma}

\def\beq{\begin{eqnarray}}
\def\eeq{\end{eqnarray}}

\def\uGeVct{\mbox{GeV}/\mbox{c}^2}

\def\numu{\nu_\mu}
\def\numubar{\bar{\nu}_\mu}

\def\piz{\pi^\circ}
\def\pip{\pi^{+}}
\def\pim{\pi^{-}}

\title{First look at the PYTHIA8 hadronization program for neutrino interaction generators}

\author{Teppei \textsc{Katori},
        Pierre \textsc{Lasorak},
        Shivesh \textsc{Mandalia}, and
        Ryan \textsc{Terri}} 

\inst{School of Physics and Astronomy, Queen Mary University of London, London E1 4NS, UK}

\email{t.katori@qmul.ac.uk}
\recdate{December 31, 2015}

\abst{Current and future neutrino oscillation experiments utilize information of hadronic final states to
  improve sensitivities on oscillation parameters measurements. 
  Among the physics of hadronic systems in neutrino interactions,
  the hadronization model controls multiplicities and kinematics of final state hadrons from
  the primary interaction vertex.
  For relatively high invariant mass events,
  neutrino interaction generators rely on the PYTHIA6 hadronization program. 
  Here, we show a possible improvement of this process in neutrino event generators,
  by utilizing expertise from the HERMES experiment.
  Next, we discuss the possibility to implement the PYTHIA8 program in neutrino interaction generators,
  including GENIE and NEUT.
  Finally, we show preliminary comparisons of PYTHIA8 predictions with 
  neutrino hadron multiplicity data from bubble chamber experiments
  within the GENIE hadronization validation tool.}

\kword{neutrino interaction, neutrino cross section, hadronization, PYTHIA, GENIE, NEUT}

\begin{document}
\maketitle

\section{Introduction, future long baseline neutrino oscillation experiments}

Current and future long baseline oscillation experiments,
such as T2K~\cite{T2K_osc}, NOvA~\cite{NuInt15_NOvA}, 
PINGU~\cite{PINGU}, ORCA~\cite{ORCA}, INO~\cite{INO}, 
Hyper-Kamiokande~\cite{HyperK}, and DUNE~\cite{LBNE}
have neutrino energies in the 1 to 10 GeV energy region.
In the past, lepton kinematics were sufficient to extract oscillation parameters,
however, higher precision measurements can be possible by exploiting information from hadronic systems.
For example, Ribordy and Smirnov pointed out that the charge separation of atmospheric neutrinos
through the inelasticity measurement (which can be estimated from hadronic energy deposit), 
improves the PINGU and ORCA neutrino mass ordering (NMO) measurement sensitivity~\cite{PINGU_inel}.
For such a measurement, it is necessary to have correct models to describe the hadronic system,
not only differential cross-sections, but also hadronization and final state interactions of hadrons.

The neutrino hadronization problem is the topic of this article. 
First, we introduce the PYTHIA and GENIE programs,
then we show how to improve PYTHIA6 for relevant neutrino experiments
through the GENIE hadronization validation tool.
Finally, we discuss the possible transition from PYTHIA6 to PYTHIA8 for
GENIE and NEUT neutrino interaction generators, including our preliminary results. 

\section{GENIE AGKY hadronization model\label{sec:genie}}

\begin{figure}[tb]
\includegraphics[width=1.0\textwidth]{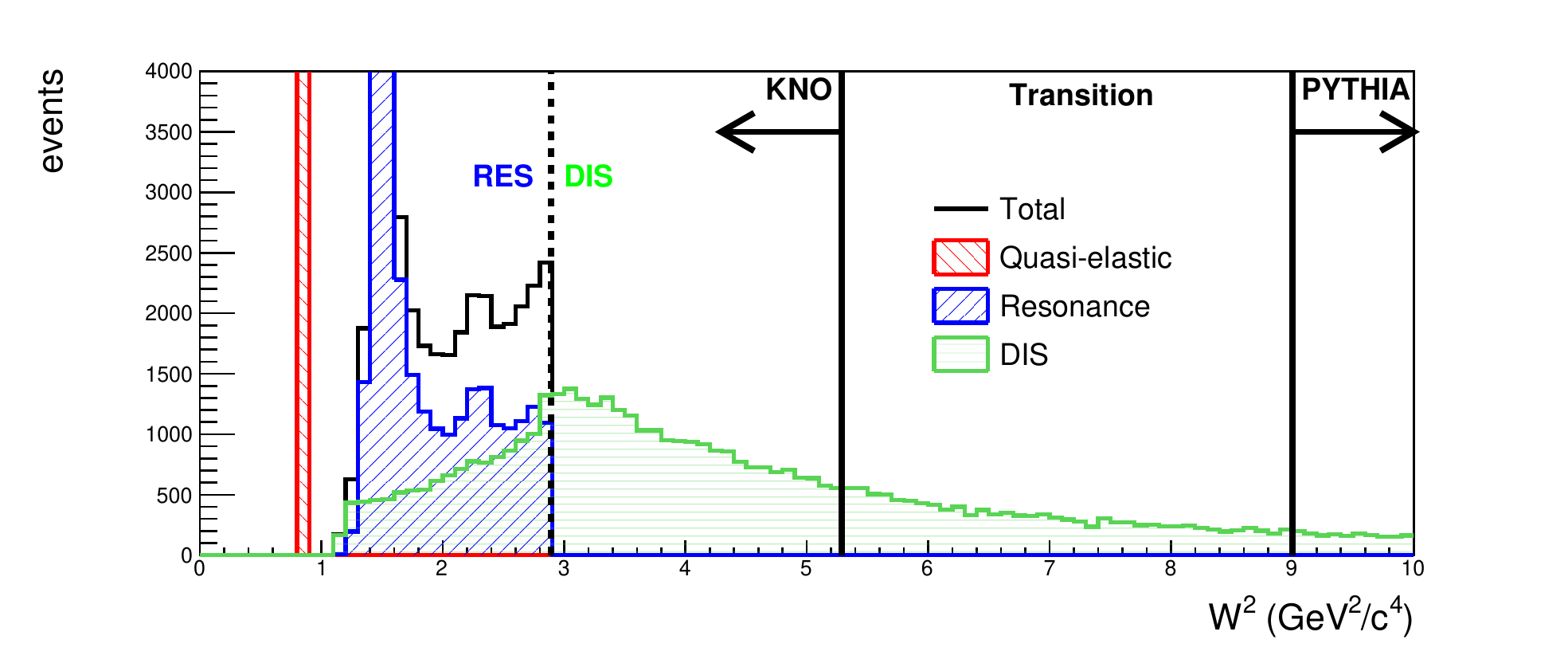}
\caption{\label{fig:KNOtoPYTHIA}
(color online) 
$W^2$ distribution of $\numu$-water target interaction in GENIE. 
For the flux, we use an atmospheric $\numu$ neutrino spectrum at the South Pole~\cite{Honda3}. 
Left red hatched region is quasi-elastic scattering, 
middle hatched region is resonance interactions, 
and right green hatched region is from DIS.  
The $W$ distribution can be split to three regions, 
KNO scaling-based model only region, 
PYTHIA only region, and the transition region.
}
\end{figure}

GENIE is a C++ based neutrino interaction MC generator~\cite{GENIE}. 
In the few-GeV energy region which are particularly important in oscillation experiments,
GENIE employs the AGKY hadronization model ~\cite{AGKY,Tingjun}.

The AGKY model is split into two parts.
At lower energy regions,
a phenomenological description based on the Koba-Nielson-Olesen (KNO) scaling law is used~\cite{KNO}.
First, averaged charged hadron multiplicity data are fit
to a function of invariant mass squared, $W^2$,
in order to extract the parameters $a_{ch}$ and $b_{ch}$, 
\beq
\left<n_{ch}\right>=a_{ch}+b_{ch}\cdot logW^2~,~
\label{eq:abW2}
\eeq
then, the total averaged hadron multiplicity is deduced
to be $\left<n_{tot}\right>=1.5\left<n_{ch}\right>$.
In this way, averaged hadron multiplicity is assigned for any interaction. 
To simulate the actual hadron multiplicity for each interaction, the KNO scaling law is used. 
The KNO scaling law relates the dispersion of hadron multiplicity at different invariant masses 
with a universal scaling function $f(n/\left< n\right>)$,
\beq
 \left< n \right> \times P(n) = f\left(\frac{n}{\left< n\right>}\right)
\label{eq:KNO}
 \eeq
where $\left<n\right>$ is the averaged hadron multiplicity and $P(n)$ is the probability of generating $n$ hadrons. 
The scaling function is parameterized by the Levy function,
\beq
L(z,c)= \frac{2 e^{-c}c^{cz + 1}}{\Ga(cz+1)}~,~
\label{eq:Levy}
\eeq
$z=n/\left< n\right>$ and an input parameter $c$. 
The input parameter is used to tune the function to agree with data,
which is mainly taken from the Fermilab 15-foot bubble chamber experiment~\cite{Zieminska}.

At higher energy regions the AGKY model gradually transitions from the KNO scaling-based model to PYTHIA6 discussed later. 
A transition window based on the value of the invariant hadronic mass $W$ is used, 
over which the fraction of events hadronized using the PYTHIA(KNO) model increases(decreases) linearly.
The default values used in the AGKY model are 
\beq
&1.& W<2.3~\uGeVct, \mbox{KNO scaling-based model only region,} \label{eq:w1}\\
&2.& 2.3~\uGeVct<W<3.0~\uGeVct, \mbox{transition region, and} \label{eq:w2}\\
&3.& 3.0~\uGeVct<W, \mbox{PYTHIA6 only region.}\label{eq:w3}
\eeq

More specifically, in the transition region, for an interaction with W,
the probability to choose the PYTHIA hadronization model is given by $\frac{W-2.3}{3.0-2.3}$. 
Fig.~\ref{fig:KNOtoPYTHIA} graphically shows this situation. 
This is the $W^2$ distribution for $\numu$-water interactions simulated with GENIE. 
Here, we used the atmospheric $\numu$ neutrino spectrum at the South Pole~\cite{Honda3}. 
As one can see, the $W^2$-distribution in this energy region can be split into three main interaction modes, 
quasi-elastic (red hatched, left peak), resonance (blue hatched, middle), 
and DIS (green hatched, right).
The AGKY model is applied to DIS interactions.  
Also, note that the DIS is extended to the low $W^2$ region to describe 
non-resonant interactions in the resonance region.

\vspace{0.5cm}
All studies in this paper use GENIE version 2.8.0. 
Figiures~\ref{fig:cMulCh} and \ref{fig:cMulCh8} are
generated by the hadronization validation tool in GENIE,
originally developed by Costas Andreopoulos (Liverpool/STFC) and Tinjun Yang (Fermilab).

\section{PYTHIA6, the standard hadronization model for neutrino interaction generators\label{sec:pythia}}

\begin{figure}[tb]
  \includegraphics[width=0.5\textwidth]{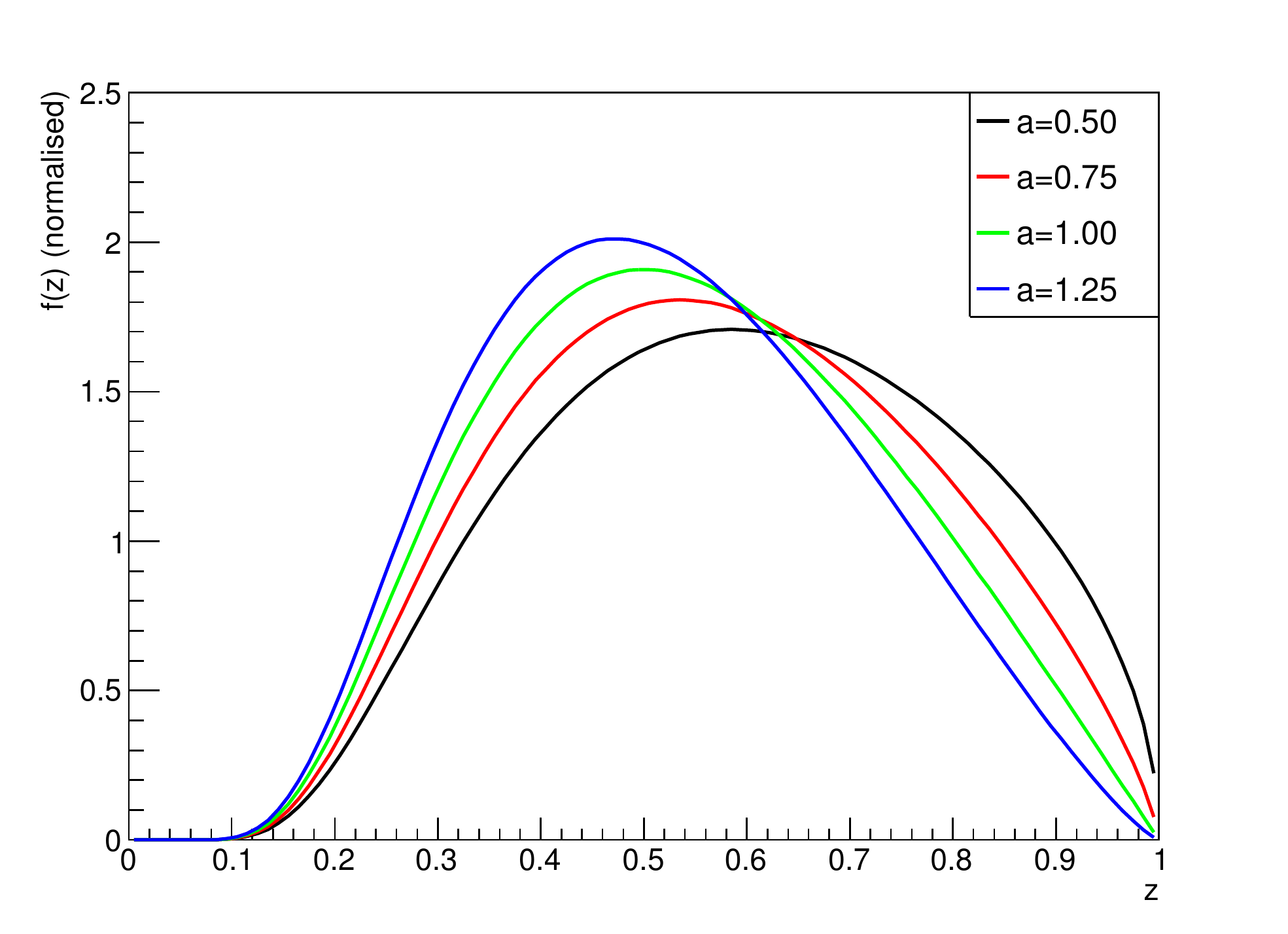}
  \includegraphics[width=0.5\textwidth]{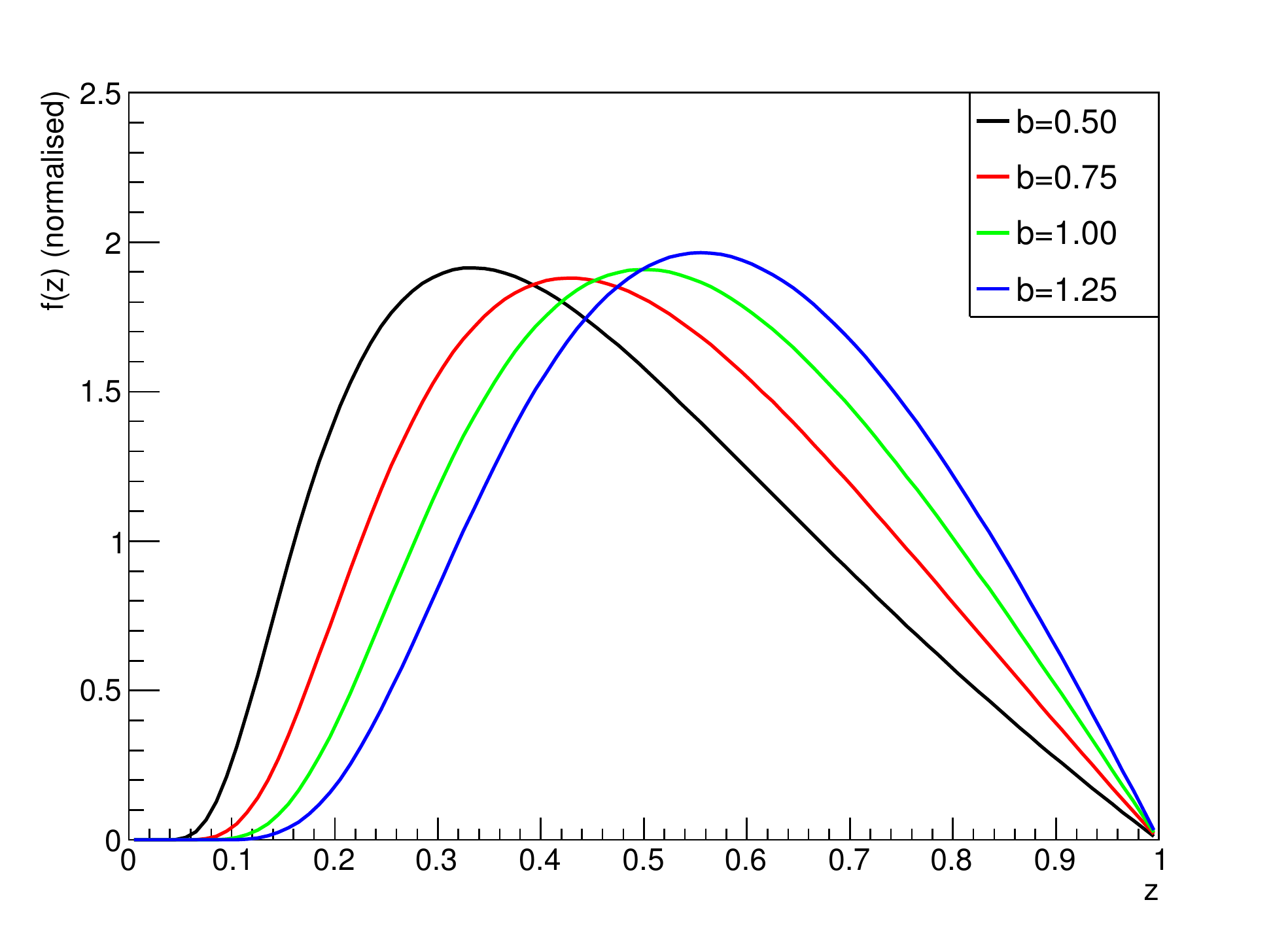}
\caption{\label{fig:lundfunc}
  (color online)
  The Lund symmetric fragmentation function
  (Eq.~\ref{eq:lundff}) for different values of
  Lund $a$ and Lund $b$. The parameter is altered while keeping all
  other variables fixed~\cite{Teppei_had}.
}
\end{figure}

The PYTHIA Monte Carlo (MC) generator~\cite{PYTHIA6,PYTHIA8}
is one of the standard hadronization tool for high energy physics experiments.
Specifically the neutrino interaction generators, GENIE~\cite{GENIE}, NEUT~\cite{NEUT}, NuWro~\cite{NuWro}, and GiBUU~\cite{GiBUU},
all rely on PYTHIA6 hadronization program~\cite{PYTHIA6} at relatively higher $W$ region.
Fragmentation in PYTHIA is described by the Lund string fragmentation model, 
which is a model based on the dynamics of one-dimensional 
relativistic strings that are stretched between colored partons. 
The hadronization process is described by break-ups in the strings through 
the production of a new quark-antiquark pairs. 
The fraction of $E+p_z$ taken by the produced hadron is given by the variable $z$, 
defined by the hadron energy $E$ and energy transfer $\nu$ ($z=E/\nu$). 
An associated fragmentation function $f(z)$ gives the probability that a given $z$ is chosen. 
The simplified Lund symmetric fragmentation function is given by,
\beq 
f(z)\propto z^{-1}(1-z)^{a}\cdot exp\left(\frac{-bm_\perp^2}{z}\right)~.
\label{eq:lundff}
\eeq
Here, $m_\perp^2$ is the transverse mass of the hadron ($m_\perp^2\equiv m^2+p_\perp^2$). 
The Gaussian term describes quantum tunneling in the transverse direction, 
and tunable  ``Lund $a$'' and ``Lund $b$'' parameters decide the longitudinal distribution of energy. 
Thus, these two parameters mainly decide how to distribute available energy to the produced hadrons. 
Fig.~\ref{fig:lundfunc} shows the Lund symmetric function~\cite{Teppei_had}. 
Larger Lund $a$ and smaller Lund $b$ parameters shift
the fragmentation function to a lower $z$ region. 
The values of these parameters are obtained from the shapes of the measured fragmentation functions, 
and default values of Lund $a$ and Lund $b$ in PYTHIA6.3 are 0.3 and 0.58~$\uGeVct$, respectively.

\section{Averaged charged hadron multiplicity\label{sec:avmulch}}

\begin{figure}[tb]
  \includegraphics[height=0.25\textheight]{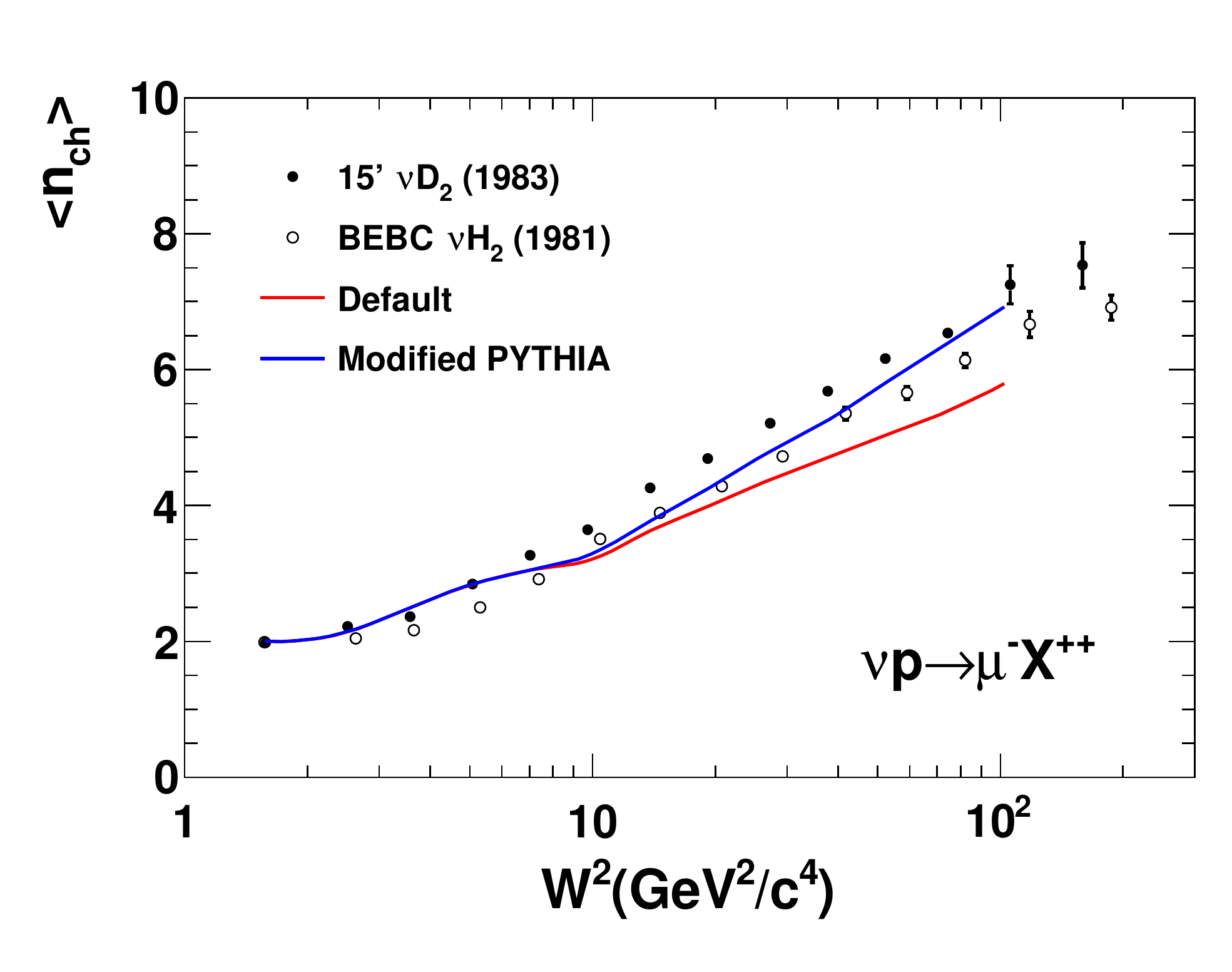}
  \includegraphics[height=0.25\textheight]{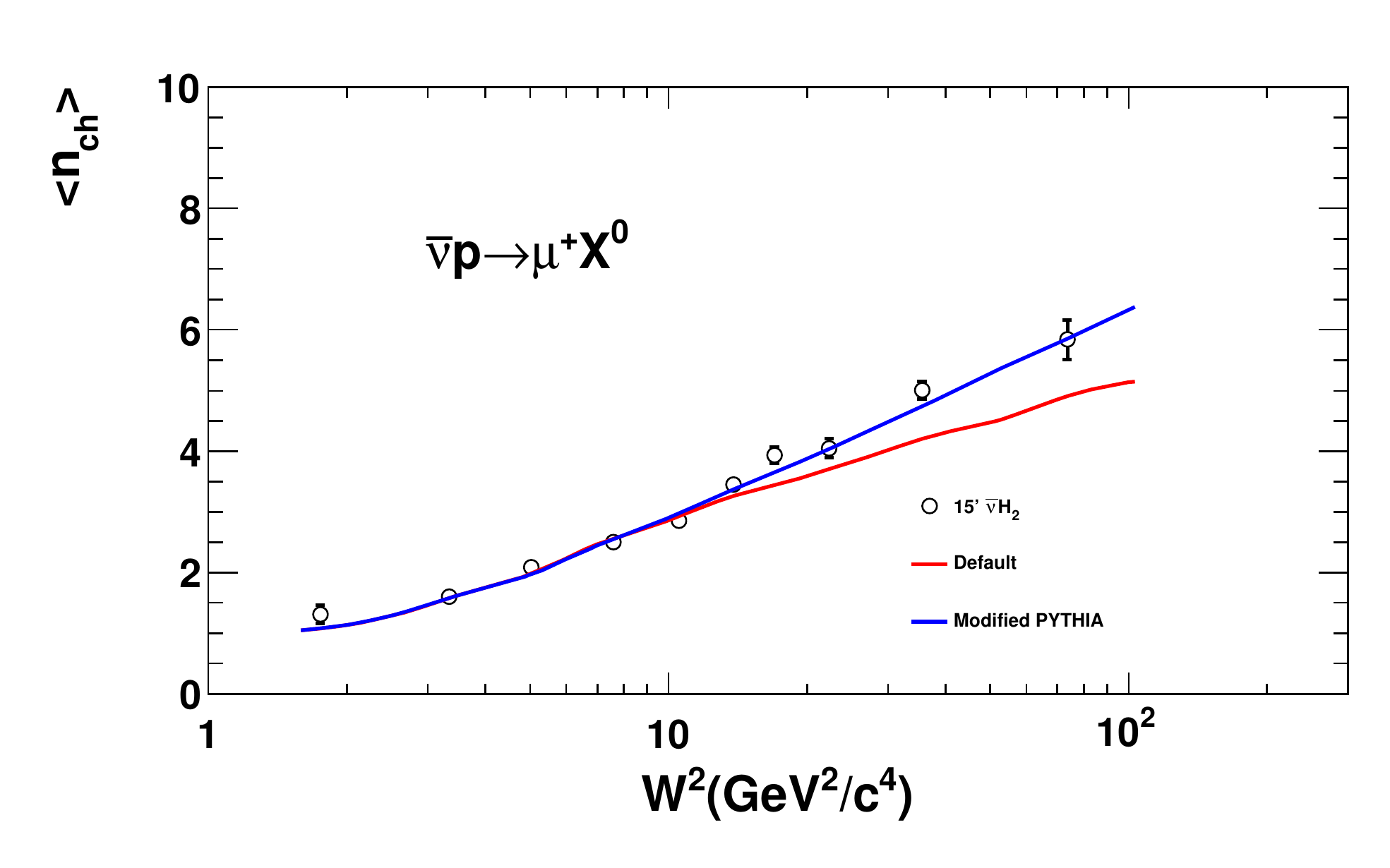}  
\caption{\label{fig:cMulCh}
(color online) 
Averaged charged hadron multiplicity plots~\cite{Teppei_had}. 
Here, two predictions from GENIE are compared with bubble chamber neutrino hadron production data
from $\numu-p$ (left) and $\numubar-p$ (right) interactions~\cite{Zieminska,BEBC_D2,Derrick}.
The lower black line is from default GENIE PYTHIA6, and the upper curve is PYTHIA6 with tuned PYTHIA6.
}
\end{figure}

Averaged charged hadron multiplicity data is fundamental in the development of hadronization models. 
It describes the average number of charged hadrons, mainly $\pip$ and $\pim$, 
measured as a function of invariant hadron mass $W$. 
Neutrino hadronization models are largely guided by such data 
from bubble chamber experiments.   
Recently, Kuzmin and Naumov performed detailed study of neutrino bubble chamber data, 
and chose the best sets of data to tune their model~\cite{KuzminNaumov}. 
They also show that neutrino interaction generators, 
such as GENIE~\cite{GENIE}, NuWro~\cite{NuWro}, and GiBUU~\cite{GiBUU}, 
all appear to underestimate averaged charged hadron multiplicity.
Note, it was also shown that the NEUT neutrino interaction generator~\cite{NEUT}
underestimates averaged charged hadron multiplicity~\cite{Connolly}. 
This problem originates from the inadequate default setting of
PYTHIA parameters for low energy experiments, 
and as we show below, this can be fixed easily by tuning few PYTHIA parameters.  

Fig.~\ref{fig:cMulCh} shows the data-MC comparison of the averaged charged hadron multiplicities
in $\numu-p$ and $\numubar-p$ interactions~\cite{Teppei_had}. 
Here, the two curves represent predictions from default GENIE and 
GENIE with a PYTHIA6 tuned using one of the parameter set from the HERMES experiment~\cite{Rubin:2009zz}. 
Note, because the $W<2.3~\uGeVct$ range of the AGKY model is hadronized using the KNO scaling-based model,
these two curves should be identical at $W<2.3~\uGeVct$. 
As you can see, the HERMES tune describes the data better.
This is because, in general, HERMES parameterization increases the averaged charged hadron multiplicity, 
which improves the agreement with averaged charged hadron multiplicity data from neutrino bubble chamber experiments.

The main effect of this new parameterization originates from 
the increase of the Lund $a$ parameter and the decrease of the Lund $b$ parameter (Eq.~\ref{eq:lundff}). 
As shown in Fig.~\ref{fig:lundfunc}, these changes make the fragmentation function softer. 
This enhances emissions of soft hadrons, {\it i.e.}, 
this increases averaged charged hadron multiplicity and thus it agrees better with data. 
Since neutrino experiments concerned here 
are relatively lower energy than LEP experiments
which are used to tune the default Lund-$a$ and Lund-$b$ paramerters in PYTHIA6, 
we shift the peak of the fragmentation function to
a lower $z$ value by increasing (decreasing) the Lund $a$ ($b$) parameter, respectively. 
In fact, all parameterization schemes from HERMES we checked
have a high Lund $a$ and low Lund $b$ parameter~\cite{Menden:2001gy,Rubin:2009zz,Hillenbrand:2005ke,HERMES_GluPol}. 
However, the neutrino bubble chamber hadronization data prefer a relatively smaller 
Lund $a$ parameter than most HERMES parameter sets, 
yet bigger than the default PYTHIA choice.  

\vspace{0.5cm}
More details of PYTHIA6 tuning work is described in Ref.~\cite{Teppei_had}.
\vspace{0.5cm}

Although averaged charged hadron multiplicity can be fixed by modifying shape of the Lund fragmentation function,
this prescription cannot fix all data-MC disagreements,
most notably averaged $\piz$ multiplicity and topological cross sections.

The former is tied with averaged charged hadron multiplicity through isospin symmetries. 
The default PYTHIA6 model can reproduce bubble chamber $\pi^\circ$ data correctly~\cite{Teppei_had},
this means the modified PYTHIA6 will overestimate the averaged $\piz$ multiplicity.
This situation is in contrast with what HERMES found, 
because PYTHIA6 with a suitable parameterization shows excellent agreement 
in both charged and neutral pion fragmentation functions with HERMES data~\cite{HERMES_MulCh,Joosten}.

The later problem originates with the dispersion of multiplicity predicted by PYTHIA6.
Although averaged hadron multiplicities can be modified by shifting the Lund fragmentation function,
this has only a minor impact on the width of the multiplicity distribution. 
On the other hand, at the lower $W$ region, the AGKY model can reproduce the data of multiplicity dispersions
by utilizing the KNO scaling law.
This creates a discontinuity in the multiplicity dispersions with function of $W$ in topological cross sections. 

It is not trivial to find a unique parameter set which works for all neutrino experiments.
Our goal here is to describe neutrino experiments in the 1 to 10 GeV region. 
In this conference, the NOMAD collaboration presented a new tuning scheme for PYTHIA6~\cite{NuInt15_NOMAD}.
The preliminary results show that NOMAD data prefer a higher Lund $a$ parameter and a higher Lund $b$ parameter.
However, in HERMES, when the Lund $a$ parameter is increased, the Lund $b$ parameter is always decreased. 
All in all, PYTHIA tuning for modern neutrino experiments is a developing field and clearly further studies are needed. 

\section{PYTHIA8}

PYTHIA8~\cite{PYTHIA8} is the latest version of PYTHIA for LHC experiments and is actively supported by its authors,
and it is widely used with proper tuning~\cite{Monash2013}.
There is a number of new features which have been introduced,
however, several features are removed or not implemented yet.
Here, we list three notable changes;

\subsubsection*{i. PYTHIA8 is based on C++ whereas PYTHIA6 is based on FORTRAN}
This allows PYTHIA8 to have a directly interface
with other modern C++ based software, such as ROOT and GENIE. 

\subsubsection*{ii. The default setting is suitable for LHC physics}

The primary purpose for PYTHIA8 is to make predictions for LHC physics,
and the default setting is suitable for LHC physics. 
People working on lower energy experiments are hesitant to transfer to PYTHIA8 from PYTHIA6,
however, as we discussed, the default setting of PYTHIA6 is also based on high energy experiments
and without tuning it is not suitable for 1 to 10 GeV neutrino physics. 
In other words, it is always important to tune of PYTHIA for use in neutrino interaction experiments.
Furthermore, at the low-invariant mass region, 
the basic assumptions of PYTHIA are broken and other hadronization models 
such as the KNO-scaling based model in the AGKY model are required.

\subsubsection*{iii. It does not simulate lepton-nucleon interactions}

Currently PYTHIA8 does not support lepton-nucleon scattering. 
This is a problem for NEUT since it currently relies on the PYTHIA lepton scattering library. 

\subsection*{Preliminary results of PYTHIA8 in GENIE v2.8.0}

\begin{figure}[tb]
  \includegraphics[height=0.25\textheight]{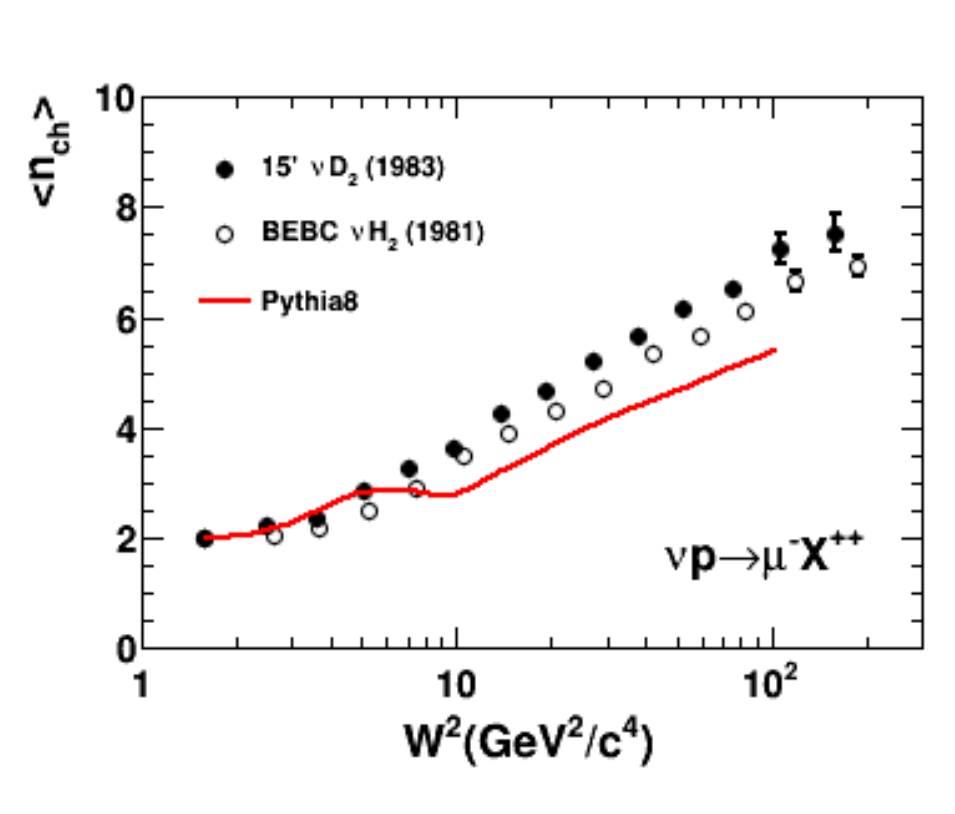}
  \includegraphics[height=0.25\textheight]{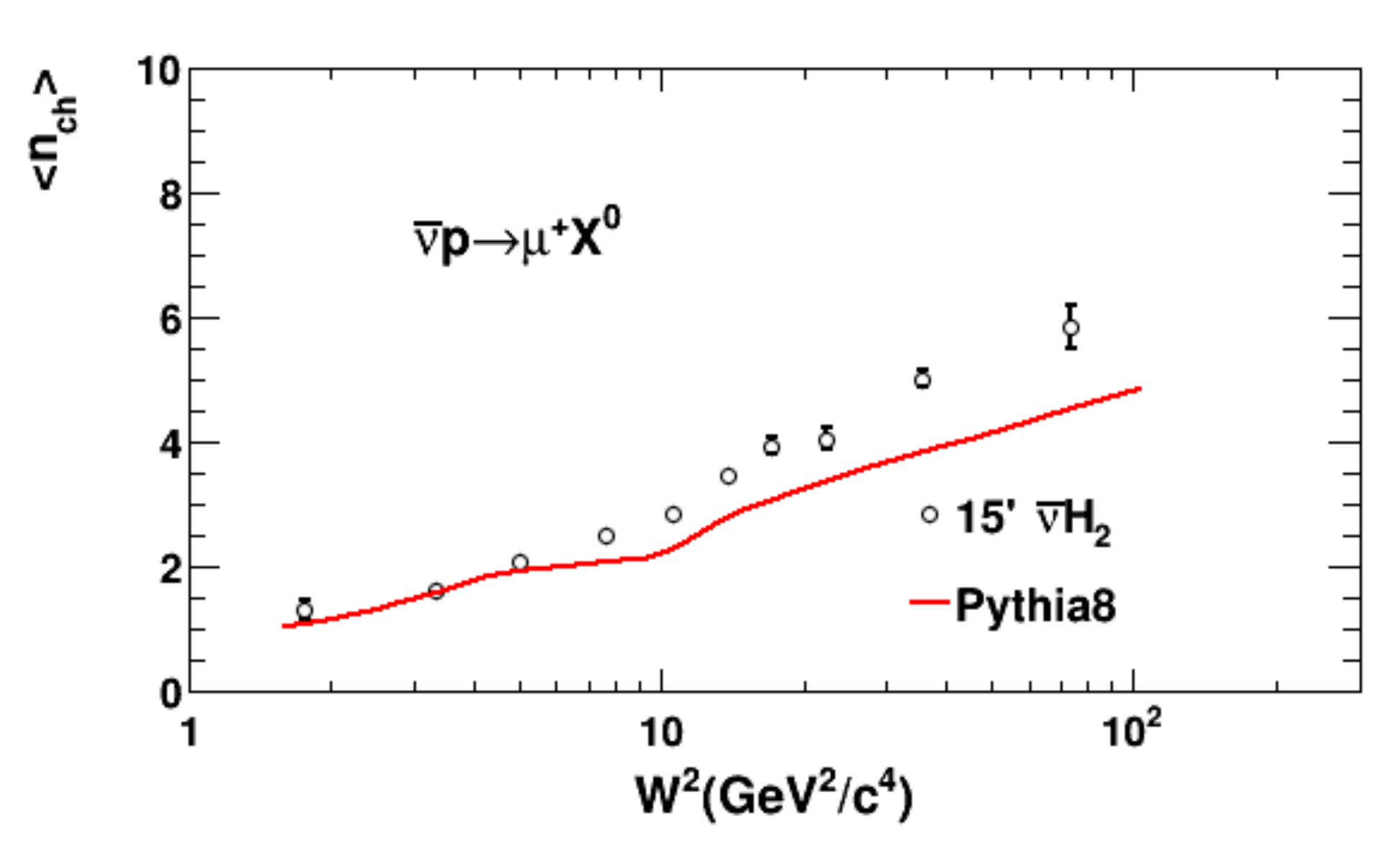}
\caption{\label{fig:cMulCh8}
(color online) 
Averaged charged hadron multiplicity plot. 
Here, the predictions from GENIE with PYTHIA8 are compared with bubble chamber $\numu-p$ 
and $\numubar-p$ hadron production data~\cite{Zieminska,BEBC_D2,Derrick}. 
}
\end{figure}

Fig.~\ref{fig:cMulCh8} shows the our preliminary results of default PYTHIA8 in GENIE. 
Here, PYTHIA6 is replaced with PYTHIA8 while keeping the same transition windows (Eqs.~\ref{eq:w1}-~\ref{eq:w3})
in the AGKY model (Fig.~\ref{fig:KNOtoPYTHIA}).
At low $W$, the predicted curves follow data points because
the hadronization process is dominated by the KNO scaling-based model.   
But as $W$ increases,
the predicted curves show large deviations with bubble chamber data.
The same feature, although less pronounced,
is observed from the default PYTHIA6 figure (Fig.~\ref{fig:cMulCh}, lower curves),
and we believe that the default parameter settings of PYTHIA8,
which are suitable for higher energy experiments such as LHC experiments,
may be the source of these large discrepancies.
However, as we demonstrated,
this can be improved by shifting the Lund fragmentation function (Eq.~\ref{eq:lundff}, shown in Fig.~\ref{fig:lundfunc}).
We also checked other distributions ($x_F$, etc),
but we do not find any unknown types of data-simulation discrepancies.  
In summary, we do not find any show stoppers for using PYTHIA8 for 1 to 10 GeV neutrino experiments
within the GENIE hadronization validation tool. 
The full PYTHIA8 implementation is ongoing. 
Once this is finished, we will start to tune to a suitable hadronization model in our energy region.

\subsection*{PYTHIA8 in NEUT}

NEUT~\cite{NEUT} is the neutrino interaction generator used by T2K~\cite{T2K_osc},
Super-Kamiokande~\cite{SuperK}, and Hyper-Kamiokande~\cite{HyperK}.  
NEUT is a FORTRAN based program,
and this clearly causes issues when using PYTHIA8 as it is based on C++. 
This problem can be solved by establishing a proper interface
between PYTHIA8 C++ functions and NEUT FORTRAN code. 
The second problem is that NEUT currently relies on the neutrino scattering program within PYTHIA,
which does not exist in PYTHIA8.
For this, a GENIE-like interface must be introduced in NEUT, 
namely the neutrino interaction scattering with the target nucleon
must happen within NEUT and then these fragments must be passed to PYTHIA8.
Once these interfaces are established,
we are ready to use PYTHIA8 within NEUT and we can begin tuning it. 
As you see, implementing PYTHIA8 in NEUT is slightly more involved than GENIE.

\section{Conclusions}

In these proceedings, we briefly discuss the usage of PYTHIA8 in neutrino interaction generators.
We found the default setting of PYTHIA8 may be optimized for even higher energy experiments
when compared with the default setting of PYTHIA6, however,
as we demonstrated this can be fixed easily by tuning the PYTHIA parameters. 
Implementation of all features of PYTHIA8 in GENIE is ongoing work,
and we expect to tune it using our expertise on PYTHIA6. 
The PYTHIA8 implementation in NEUT requires more steps than GENIE
due to the absence of lepton-nucleon scattering models inside PYTHIA8.

\subsection*{Acknowledgments}
TK thanks Ulrich Mosel for introducing this subject to us.
We thank Elke Aschenauer and Josh Rubin for useful information about the HERMES experiment,  
we thank Steve Mrenna for useful information about PYTHIA. 
We also appreciate the various help given to us by Gabe Perdue and Julia Yarba on the GENIE simulation.
Finally, TK would like to thank the organizers of NuInt15 
for the hospitality during my stay at Osaka, Japan,
and for giving the opportunity for me to present this work.

\bibliographystyle{iopart-num}
\bibliography{nuint15}
\end{document}